\documentclass[twocolumn,aps,rmp,twocolumn,a4]{revtex4}
\usepackage[latin1]{inputenc}
\usepackage{geometry}
\makeatletter

\providecommand{\LyX}{L\kern-.1667em\lower.25em\hbox{Y}\kern-.125emX\@}
\newcommand{\noun}[1]{\textsc{#1}}

\usepackage{pslatex,epsfig,floatflt}
\bibpunct{[}{]}{;}{,}{,}

\makeatother

\begin{document}

\title{\huge{Quantum complementarity, erasers and photons}}

\author{A. F. Kracklauer}

\address{Bauhaus Universit\"at, Weimar, Germany}

\begin{abstract}
Optical experiments designed to explore quantum complementarity are reanalyzed.
It is argued that, for each, a classical explanation is not only possible, but
more coherent and less contrived. The final conclusion is that these experiments
actually constitute support for criticism of the photon paradigm of electric
charged particle interaction. They offer little or nothing to say about quantum
complementarity once the photon concept is not imposed by mandate.

\homepage{http://www.nonloco-physics.000freehosting.com}
\end{abstract}
\maketitle

\section{Background}

Complementarity has become a catch-all concept to cover the weirdness of Quantum
Mechanics (QM). Mostly, this is because its originator, \noun{Bohr,} actually
proposed this notion in order to capture exactly these weird features and legitimize
them. It is arguably the case, however, that having been unable to remove what
were at first recognized as antinomies, they were just redubbed as deep, albeit
preternatural, holistic insights into atomic scale ontology. Following the popularization
of the term in Physics, \noun{Bohr} extended the concept to arenas outside
the customary boundaries of science even, until it became for him a universal
precept, the foundation of a total \emph{Weltanschauung.} History seems to show
that he was only marginally successful at promulgating philosophy. None of his
students, nor readers of his papers, agree completely on just whatever it was
he tried to convey under the term ``complementarity,'' even when restricted
to within Physics.\cite{1}

Nevertheless, through the decades, authors of textbooks and other secondary
literature have focused the notion of complementarity on the issue of particle-wave
identity. Out of this literature over time a consensus has crystallized: complementarity
refers to that feature of quantum theory to the effect that ensembles of all
entities, depending on the scale of the interaction or measurement scheme, exhibit
alternately either the properties of waves or particles. 

This has led some to suggest that there exists a single, fundamental category,
instead of particles and waves, namely ``wavicles.'' Such abstract unification,
however, does not overcome the reality of the matter, namely that this category
has two obvious components: those entities which classically are particles and
those that are waves. This division is obvious in terms of the characteristic
of mass; classical waves can be ascribed only a mass equivalent of energy. 

Through the years, various proposals have been made to render this ``schizoid''
phenomena less mysterious. For each of the two subcatagories different conceptions
have been proposed; the wave-like behavior of particle beams, for example, has
been attributed to various forms of `pilot waves.' As is well known, \noun{de
Broglie,} who first suggested that particles should exhibit wave phenomena,
more or less explicitly envisioned that particles were a sort of material kernel
embedded in such a pilot wave.\cite{2} The wave portion was credited with navigating
for the kernel without itself alone interacting with measuring devices, a behavior
that eventually led to coining the term ``empty wave.''\footnote{%
Just when the term ``empty wave'' was introduced is unclear to this writer.
As the idea is derived from \noun{de Broglie}'s pilot wave, many empty wave
properties have been under consideration for a long time. See \noun{Selleri}\cite{3}
for a recent discussion.
} Proponents of this notion seem to have abandoned it, perhaps because of the
internally contradictory conception that a pilot wave can react to all material
boundaries except those used in connection with measurements. How can any wave
know the difference in purpose of any particular material object it encounters?
When is it supposed to communicate with objects so that it can navigate, and
when is it to ignore them as the intervention of an experimenter? 

This writer has proposed an extention of the pilot wave idea to the effect that
such pilot waves are classical electromagnetic background waves attached to
particles by virtue of their motion through the random electromagnetic background
radiation that is hypothesized as the basis of a theory known as `Stochastic
Electrodynamics' (SED), a theory which seeks to rationalize quantum phenomena.\cite{4}
Whether this effort is successful, the reader is invited to pursue independently;
it is not the central theme of this paper, which concerns the other subcategory,
the particle-like behavior of electromagnetic waves.

Particle-like behavior of waves is nowadays credited mostly to \noun{Einstein}
in connection with the Photoelectric Effect. Actually, however, \noun{Einstein}
proposed only that energy transfer occurs in `packets,' it was later that the
idea, that the incoming radiation itself is somehow subdivided, captured the
common imagination. Still, even today, the conception is vague. Sometimes considered
a `mode' of the vacuum, and thereby present throughout the universe on the one
hand, elsewhere such packets are envisioned as ``needle radiation'' with small
lateral cross-section, because absorption seems to be point-like. On other occasions
they are thought to be shockwave-like, because absorption can be rapid, seemingly
instantaneous. In combination, such solutions are difficult to associate with
the inhomogeneous wave equation (with sources), as derived from \noun{Maxwell}'s
equations. 

Additional support for packaged radiation was found in the scattering of electromagnetic
radiation from electrons, i.e., the \noun{Compton} Effect. This effect is,
in a certain elementary manner, just a particular application of the photoelectric
effect. In any case, the notion of packages, as a model for waves, particularly
electromagnetic waves, has become virtually universal nowadays as the `photon'
paradigm. No modern commentator fails to note that, fundamentally, light is
a stream of photons without branding himself an ignoramus. This, in spite of
the spectacular applicability of non quantum \noun{Maxwell} theory. It is
interesting to speculate on just why this is so; this writer likes to suggest
that the cause is that electromagnetic waves, as such, are never detected. In
fact the mechanism of the interaction of charged particles is completely beyond
the reach of experimental science, which is restricted to `photocurrents,' for
which there are two variants: those acting as sources in antennas, and those
driven by the consequences of the former on the other end in detectors.\footnote{%
Actually, every charge \emph{vis-\`a-vis} all others, plays both roles; but,
to date there is no widely accepted closed form of electromechanics taking this
into account.\cite{4a}
} In spite of the vividness of the images from paradigms of what transpires in
between, either as electromagnetic waves or photon streams, all effective experience
for mortals is actually limited to these currents, information on which is then
used mentally to \emph{infer} just how these two currents have interacted in
detail across space and through time. This understanding makes photo detection
the crux of the matter, and arguably, the essential reason for the utility of
the ``photon'' concept.

Here we come to the point of this paper. The digitization, or `quantization'
if one prefers, of charges, is \emph{not} a result of quantum theory, but an
independent empirical fact that serves as an hypothetical input into both classical
and quantum theories of physical phenomena; thus, the ``quantum'' behavior of
electromagnetic waves as ensembles of particles, may not be at all ``quantum,''
but simply a consequence of the discreteness of the charges making up the material
of detectors. If we take this observation as a scientific proposition to be
tested for internal (theoretical) consistency and external (empirical) verification,
then it cannot be regarded as anything other than just another task for work-a-day
science. In this spirit, it shall be addressed herein.

\section{Prototypical wave detectors}

Electromagnetic wave detectors come in a variety of constructions, each intended
for a particular frequency range. In those lower portions of the electromagnetic
spectrum where current technology enables following time development of electromagnetic
fields, there is little of special interest for fundamental quantum physics
analysis. It is only in that range in which this is no longer possible that
quantum phenomena seem to arise.\cite{5} Detectors in this range do not follow
the undulatory time development of incoming radiation. Instead, they absorb
an undetermined number of cycles, and respond then with some bulk transition. 

Photographic plates are a good example. Incoming radiation at some point triggers
a chemical reaction in molecules embedded in an emulsion that, as differentiated
compounds, then serve during development as seeds for some macroscopically visible
effect. Likewise, for the photoelectric effect. Incoming radiation eventually
results in the transition of a valence electron to the conduction band where
it is then drawn off into circuitry to be amplified and registered as the `detection
event.' The historical photoelectric effect considered by \noun{Einstein}
has an additional feature that is widely taken as symptomatic of an essential
quantum character. It is that the stimulated photocurrent appears virtually
instantaneously after the start of illumination. The argument that this is to
be explained only by assuming that the incoming energy is bundled in compact
packages, nowadays called `photons,' assumes that the electrons, before illumination,
were all resting at an energy so low that it would require a measurable duration
of exposure to the incoming radiation before they could be energized sufficiently
to enable them to overcome binding forces.\cite{6} Such an assumption is not,
however, beyond reproach. It could just as well be assumed that the electrons
in the detector mass had a distribution of energies attributable to prevalent
or incoming noise so that of the whole population, a certain portion at any
given instant has noise energy bringing them up to nearly the escape level.
Then when a coherent stimulation signal arrives, those electrons at this otherwise
temporary `almost' escaped energy level are boosted over the threshold very
quickly, virtually instantly. In fact, \noun{Lamb} and \noun{Scully} long
ago have shown with such ideas that photoelectric phenomena do not entail quantum
theory; semiclassical ideas suffice.\cite{7}

The vital point made here is: the digitization of the detector reaction to incoming
radiation need not necessarily be attributed to particle-like packaging of the
incoming stimulus, i.e., photons. All the behavior involved can be explained
fully also in terms of continuous, wave-like radiation of the classical sort
known to radio, radar and communication engineers, where the discrete aspects
of low intensity measurements are due to discreteness of detector charges. In
other words, because charges are the ineluctable intermediaries between mortals
and electromagnetic waves, their contribution to the nature of an observation
cannot be evaded.

\section{Complementarity in optical experiments}

It is the purpose here to parse the logic of some modern optics experiments
plumbing the mysteries of complementarity of electromagnetic interaction. These
are, first, two `quantum eraser experiments, and then `\noun{Afshar}'s' experiment;
all three are intended to focus specifically on the issue of delimiting complementarity
in light.

\subsection{Quantum Eraser}

There are several versions of experiments demonstrating this conception proposed
first by \noun{Scully} and \noun{Dr\"uhl} in 1982.\cite{8} The basic idea
is to elaborate on a standard two-slit experiment so as to be able to mark the
signal passing through at least one of the slits, thereby permitting the determination
of which slit the purported `photon' passes through. An extra feature in this
proposal was the notion that, with clever design, it could be possible to ``erase
information'' so as to flip the case between wave and particle, most spectacularly
after-the-fact even. As is well known, textbook orthodoxy nowadays would have
it that if the slit of passage can be determined (a particle-like property),
then no interference pattern (a wave-like property) will result. 

The difficult trick in conceiving of and performing such an experiment is to
find a method to mark a signal on passage through a slit. The original proposal
was to employ multilevel atoms in place of slits. The atoms were chosen to have,
at least in principle, the same emitted decay radiation but different final
states readable by some other measurement. This scheme should allow the experimenter
to do, even well after-the-fact, or not do the second measurement; as a consequence,
the interference pattern should appear or not appear. (The sense in which the
pattern appears or not results from the alternate data reduction calculations
feasible with the existence of different elements in the data sets, after-the-fact.
It should not be understood that some visual image of an object is made to come
and go.) We gather that exactly this setup was never realized, however.

\subsubsection{Eraser and double-slit diffraction }

\begin{figure}

\centerline{\includegraphics[width=63mm]{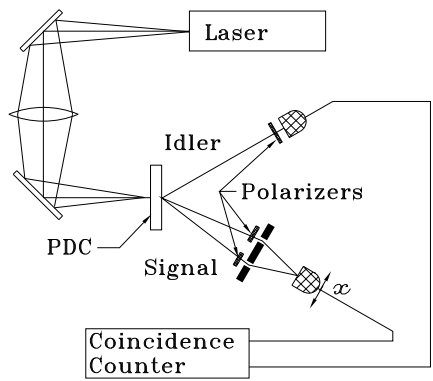}}

\caption{A setup for optically testing the `quantum eraser' effect. The basic trick is to try to use the idler emission to mark the portion of the signal passing through each of the slits in a double-slit diffraction experiment.} 

\end{figure}

A particularly clean and purely optical `quantum eraser' experiment involving
two slits, inspired by a conceptually similar proposal of \noun{Scully} et
al. \cite{8a}, was carried out by \noun{Walborn} et al. and reported in 2002.\cite{9}
The essential feature of this setup is that the beam sent towards the slits
is the signal output of a parametric down conversion crystal (PDC)\footnote{%
PDC crystals are two types: Type I in which the polarization of idler and signal
are anticorrelated; and, Type-II in which which they are correlated. 
}; the idler is directed separately through another polarizer and then to a detector.
See Fig 1. Here the marking in the slits is considered achieved by placing polarizers
before each slit. If the two polarizers before the slits are set to orthogonal
positions, then the signals impinging on the registration screen are independent
and do not form an interference pattern. This feature is thought to conform
with the complementarity principle, i.e., that which-way information provided
by a polarization tag put on the signals as they pass through the slits, is
said to destroy the wave-information revealed by the interference pattern.

\begin{figure}\centerline{\includegraphics[width=63mm]{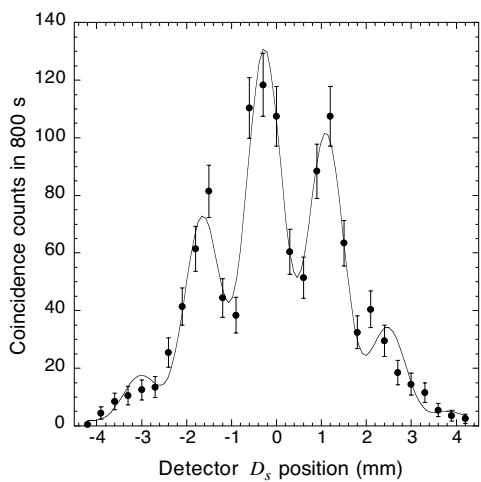}}\caption{The fringe pattern seen when the polarizer in the idler beam is vertical.\cite{9}}\end{figure}

Now, however, if coincident counts between data points taken behind the slits
with points seen from the idler beam are analyzed, then the interference effect
can be made to reappear---in the coincidence counts. That is, if the polarizer
in the idler beam (let us assume from a Type-I PDC) is set to horizontal, then
the signal in the vertical channel behind the slits will yield correlated ``hits,''
(Fig. 2) by effectively filtering out those hits in the horizontal channel for
lack of a mate in the idler channel. If the idler polarizer is changed to vertical,
then the correlated hits are in the horizontal channel behind the slits, as
now the vertical component is filtered out (Fig. 3). If the polarizer is removed
from the idler beam, then no correlations with the slits can be made; the only
hit pattern available shows no interference, and it is said that ``which-way
information has been erased'' (Fig. 4). 

This behavior is said to illustrate quantum complementary. But does it? First
note that the various signals used are just different states of polarization.
This is very significant because polarization is a phenomenon fully explained
by \noun{Stokes} circa 50 years before the need for quantum theory was known.\cite{10}
It is a fully classical phenomenon. How then, can this phenomenon be used to
plumb a principle of quantum mechanics? The only even remotely quantum aspect
to this story is the implicit assumption that the signals are made up of `quantum'
objects, i.e., photons, so that it is presumed their behavior must be regulated
by quantum mechanics. But, as was argued above, the possibility of fully non
quantum character has not been rigorously excluded.

\begin{figure}\centerline{\includegraphics[width=63mm]{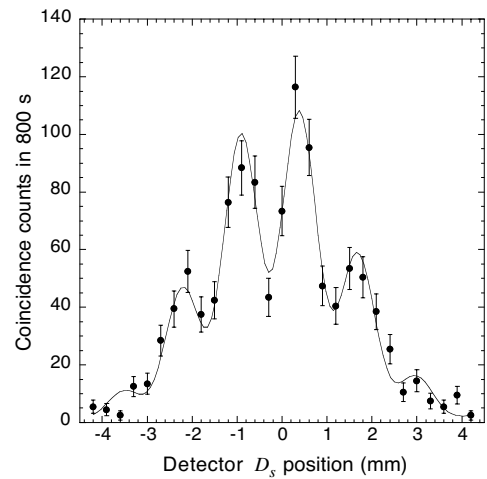}}\caption{The antifringe pattern seen when the polarizer in the idler beam is horizontal.\cite{9}}\end{figure}

Moreover, since there are two slits and two different interference patterns,
with arguments that are brief and loose (i.e., ``it is clear that \( \ldots  \)'')
in the customary interpretation, an association is made to `which-way' information.
This is done in spite of the fact that each interference pattern is indisputably
a two slit pattern. Thereafter, even while observing that the \emph{sum} of
these two patterns is what is observed when the idler polarizer is removed,
is is said that this results from \emph{erasing} some information. One ought
be excused for finding this apologia at least lexicographically pathological.

Thus, those who find such imagery based on quantum mechanics at all obscure
and contrived may be well disposed to consider the following classical rendition.
Let us take it that the signal directed towards the slits is a vertically polarized
classical electromagnetic beam. If there are no polarizer filters in front of
the slits, after passing through both slits the signal shows interference on
the registration screen. Next let us place polarizers with axes set at \( -\pi /4 \)
to the vertical before one slit, and at \( +\pi /4 \) before the other slit.
Now, each polarizer will split the incoming vertically polarized electric field
into two components one vertical and one horizontal, each with amplitude \( 1/\sqrt{2}\times 1/\sqrt{2} \)
times the `mother' signal as determined by \noun{Malus}'s Law, once as a projection
onto the polarizer axis and then again to find the component in either the vertical
or horizontal direction. In turn, each orthogonal component or `daughter' signal
will result in an interference pattern on the registration screen. But, one
pattern will be fringes (Fig. 2), while the other is antifringes (Fig. 3). Because
the horizontal diffraction pattern is the sum of opposing components, a phase
term of \( \pi  \) is inserted and the pattern comprises antifringes; in contrast
the vertical components are codirectional and in phase. These two patterns together
add up to a total pattern without interference (Fig 4). That is, if the daughter
horizontal interference pattern can be written
\begin{equation}
\label{1}
I_{h}(x)=ke^{-ax^{2}}\cos ^{2}bx
\end{equation}
 and the daughter vertical pattern\footnote{%
The phase difference of \( \pi /2 \) in these expressions for intensity equals
\( \pi  \) in \( E \)-field amplitudes.
} 
\begin{equation}
\label{2}
I_{v}(x)=ke^{-ax^{2}}\sin ^{2}bx,
\end{equation}
 then the sum
\begin{equation}
\label{3}
I_{h}(x)+I_{v}(x)=ke^{-ax^{2}}(\cos ^{2}bx+\sin ^{2}bx)=ke^{-ax^{2}},
\end{equation}
 shows no interference.

In the terminology of this explanation, it can not be said that information
has been erased, although it has been concealed, using, as it were, a kind of
secret writing. It can be refound (rendered legible) as follows. As the signal
was one output from a parametric down conversion crystal (PDC), the conjugate
idler then, depending on whether the crystal was of type I or II, will be horizontal
or vertical polarization respectively. If type I, then the vertical contribution
to the total pattern behind the slits will have nearly perfect correlated partner
pulses in the horizontal mode in the idler branch. By counting only those hits
in the signal pattern behind the slits for which there is a companion hit in
the horizontal idler mode, effectively the horizontal contribution to the signal
pattern is filtered out, revealing the vertical interference pattern. This procedure
can be called `coincidence filtering.' When such coincidences are the result
of polarization, such filtering is a purely non quantum procedure.

\begin{figure}\centerline{\includegraphics[width=63mm]{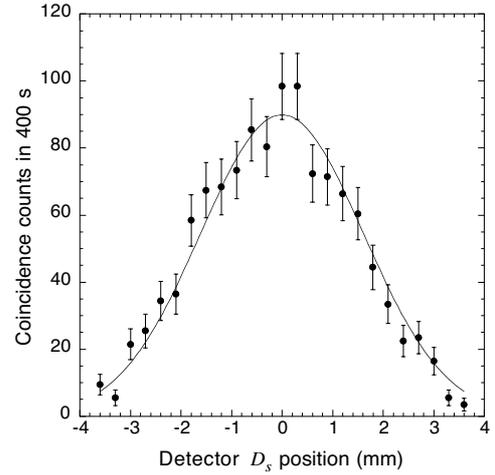}}\caption{The sum of the fringe and antifringe patterns seen when there is no polarizer in the idler beam.\cite{9}}\end{figure}

On the basis of this analysis it can be said that such a ``quantum eraser''
is neither quantum nor an eraser; as such, it has little to contribute to a
debate regarding the significance of quantum complementarity. Furthermore, the
only feature resembling quantum theory is the fact that the intensity of the
signals used is so low that detailed structure, i.e., its composition as a stream
of point charges, that is electrons, is visible as individual ``hits.''

\subsubsection{Delayed choice eraser}

Another particularly interesting `quantum eraser' experiment put the decision
to measure wave- or particle-like characteristics to the internals of the setup,
thereby taking it away from the experimenter. The experimental setup is depicted
in Fig 5. \begin{figure}\centerline{\includegraphics[width=78mm]{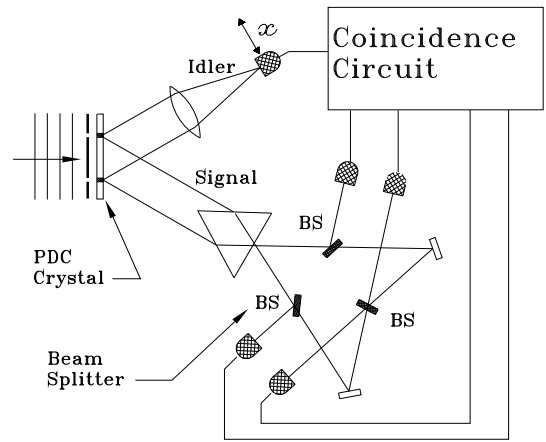}}\caption{A schematic showing the setup of the delayed choice quantum-eraser experiment. Salient features include the long optical path length of signal in relation to the idler branch, and, the use of passive beam splitters (BS) that, in effect, radomly compile ensembles that do or do not correlate with the idler branch. }\end{figure}

The salient feature in this experiment, and distinguishing it from the one just
considered, is the use of beam splitters. At the so-called single photon level
they are known to be virtually absolutist in splitting a beam by whole photons;
either a whole photon goes through, or a whole photon is reflected, not both
by dividing incoming energy. In this experiment their role is to take the decision
regarding which correlation shall be registered out of the hands of a mortal,
i.e., the experimenter, and put it in the hands of \noun{Zeus}. Then, after-the-fact,
each subensemble marked by which detector combination fired, is separately analyzed
for intensity as a function of lateral detector displacement in the idler branch.
Ensembles that correspond, again, to those made by coincidences triggered by
just one of the components in one or the other dimension, corresponding to Eqs.
(\ref{1}) or (\ref{2}), show interference.

Beam splitters operating on a beam of the intensity yielding single hits in
the detectors, show strict anticorrelated hits, which is interpreted nowadays
as reflecting the essential, ineluctable, distinct identity of photons. According
to current understanding, the random choice of transmission or reflection is
a reflection of the fundamental uncertain, statistical, i.e., ``quantum'' character
of the universe at an atomic scale. This feature might seem, therefore, to vest
this experiment with an essential quantum element. For this experiment, however,
it is immaterial how the choice is made; the phenomena under study does not
depend on the mechanism (or lack thereof) of choice. The only relevant feature
is that the separate subensembles be compiled; because only within each does
the phenomenon of interest takes place. 

The quantum character of beam splitters, although logical on its surface, is
actually not inevitable. For it to be necessary, would require that alternative
theories are logically and rigorously excluded. Insofar as non quantum alternatives
exist, e.g.:\cite{12}, the viability of the photon paradigm is just provisional.

\subsection{\noun{Afshar's} experiment}

This experiment also aims to study complementarity.\cite{12a} The basic idea
is to set up a double-slit diffraction pattern at a middle stage of an optics
train which is continued through a lens to end in a sharp image of the slits;
see: Fig. 6. Then, its conceptual novelty consists in showing that thin opaque
obstacles placed in the beam exactly at the locations of destructive interference
in the fringe pattern midway through the train have no effect on the intensity
of the slit's image at the end of the train. The image of the slits can be sharp
only if those photons, which it would seem have to have been intercepted by
the obstacles, are in fact not removed from the beam. Indeed, if one slit is
blocked, the obstacles do substantially degrade the image of the open slit.
Thus, one is led to question how to reconcile these contradictory facts.

\begin{figure}\centerline{\includegraphics[width=78mm]{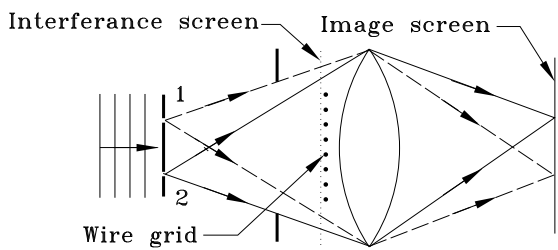}}\caption{The basic setup of \noun{Afshar}'s experiment. First the location of the nodes of the diffraction pattern on the interferance screen are determined as the location for opaque objects (wire grid, seen here on end). Then this screen is removed while leaving the grid in place. Finally the image of the slits is observed on the image screen. The presence of the grid does not affect the quality of the slit image.}\end{figure}

The very precise final image of the slits gives very precise which-way, i.e.,
particle-type information (in the sense of geometrical optics) on which slit
the light for each slit image has passed through. At the same time, knowing
just exactly where to place the opaque objects requires very precise knowledge
of the wave-optical character of the beam. Since both types of knowledge are
in play and are known to the experimenter at the same time, this knowledge is,
arguably, in violent contradiction to the complementarity principle. 

Once again, however, a central question is just what is `quantum mechanical'
in the involved phenomena? There is nothing in the device chain for which otherwise
quantum theory is required. Indeed, in \noun{Afshar}'s paper analyzing this
experiment, one does not find a single occasion for which a factor of \noun{Planck}'s
constant is required. The only quantum-appearing structure that is used is the
\noun{Born} association of the probability of presence (here of a `photon')
being proportional to the square of the field. In this case the field is actually
the electromagnetic field, so that the \noun{Born} hypothesis completely overlaps
the theory of photocurrent generation, i.e., that it is proportional to the
square of the electric field, a fact that might be seen even as the underpinning
for \noun{Born}'s interpretation. Moreover, \noun{Afshar}'s analysis calls
on the notion that properties of a light beam are considered continously variable
from wave to particle, a concept that wreaks havoc with the concept of `particle.'
Once again, there is little to nothing here addressing any feature of the complimentary
principle from \noun{Bohr.} On this basis it can be taken that this experiment
actually addresses the issue of the tenability of the `photon' paradigm. It
renders the ray-optical viewpoint of photon propagation problematical while
endorsing wave propagation. In other words, it is conceptually a variation of
the extensive experimental program of \noun{Roychoudhuri\cite{13},} who identified
a number of internal inconsistencies in photon imagery.

\section*{Conclusions}

When is a phenomenon a quantum phenomenon? This question can be deeper than
it appears at first sight. A complication arises in that some non quantum structure
from classical physics fits perfectly well within the vocabulary, notation,
algorithms and interpretation of quantum theory without, however, actually depending
on any hypothesis unique to quantum mechanics. Vectors of absolutely any sort,
for example, can be symbolized by `bras' and `kets.' Insofar as the bra-ket
notation encompasses abstract vector space structure, it can be self-consistently
used whenever this structure obtains. To this writer's mind, this leaves only
one option open for defining the category of quantum phenomena, namely: \emph{a
quantum phenomenon is one that absolutely cannot have its patterns encoded mathematically
without calling on a unique fundamental hypothetical input of Quantum Mechanics.}
In the final analysis, this means that the noncommutivity of phase (or quadrature)
space must be essential. Therefore, any phenomenon that can be explained without
reference to this noncommutivity, no matter how unappealing the non quantum
argument, is not a genuine ``quantum phenomenon,'' even when it can also be
explained with quantum mechanics.

Just here there arises a premier example of non quantum structure embedded in
the quantum formalism. The noncommutivity found in `qubit' space of polarization
space or of spin space is widely thought to be of a quantum nature. However,
the group in play here is \( SU(2) \) which is homeomorphic to \( O(3) \).
The latter encodes the structure of rotations on a sphere which is a totally
geometric structure having nothing to do with quantum mechanics. By cause of
homomorphism, if \( O(3) \) is geometric, then \( SU(2) \) must be also. The
noncommutivity uniquely evident in quantum mechanics is in phase space and is
a dynamical feature, not just geometry. 

In the experiments considered above, noncommutivity on phase space plays no
role in describing the observed phenomenon. Thus, it is concluded that `quantum'
is not an appropriate adjective. The term `eraser' is also inappropriate. To
erase, etymologically, is derived from Latin words meaning `to scrape out.'
Clearly this meaning is derived from times when writing was often carved in
stone. Carving out words is essentially different than filling in the letters
rendering them illegible (with due attention to color and texture). Competent
restoration can recover a concealed message; destroyed informations is lost
forevermore. None of this has anything to do with the depths of a complementarity
principle for electromagnetic radiation.

The key or essential feature of what heretofore goes under the rubric of
(optical) ``quantum eraser'' is the phase shift between the vertical and
horizontal components. This shift allows the two components to fully
compliment each other as seen in Eq. (\ref{3}). Because this is a simple
geometric effect, it must arise in a great variety of situations. The
euphoric bedazzlement at its implications for the nature of `time' in
Quantum Mechanics as reported for some observers in \cite{17}, from this
point of view, is a reaction to the logical jujitsu entailed in trying to
understand nature on the basis of inappropriate assumptions, in this case,
the photon paradigm.

This is not to claim that a complementarity principle has nothing to do with
quantum mechanics. Although any such principle seems to be an expression of
bandwidth considerations from classical wave theory, in fact the quantum one
is intimately connected to the \noun{Heisenberg} uncertainty relation, which,
unlike electromagnetic bandwidth, is scaled by \noun{Planck}'s constant, and
must, therefore, refer essentially to a quantum phenomenon. The pertinent question
then may be: how and why for \noun{De Broglie} waves (sometimes called `matter
waves') is bandwidth regulated in a fundamentally distinct manner from that
for electromagnetic waves? But, this issue pertains to entities that are indisputably
particulate in a classical limit, and not to continuous, undulatory radiation.

\end{document}